\documentclass{kluwer}    

\newdisplay{guess}{Conjecture}
\usepackage[dvips]{graphicx}

\begin{document}                                                                                   
\begin{article}
\begin{opening}         
\title{Searching for tidal tails in  Galactic dwarf spheroidal
satellites} 
\author{D. Mart\'\i nez-Delgado}  
\runningauthor{Mart\'\i nez-Delgado et al.}
\runningtitle{Tidal tails in Galactic dSphs}
\institute{ Instituto de Astrof\'\i sica de Canarias, 38200 La Laguna, Spain }
\author{ A. Aparicio}
\institute{ Instituto de Astrof\'\i sica de Canarias, 38200 La Laguna, Spain }

\author{ M. A. G\'omez-Flechoso}
\institute{Geneva Observatory, CH-1290 Sauverny, Switzerland}

\date{August 15, 2000}

\begin{abstract}
 We present preliminary results of a long-term  project to investigate the process of
accretion and tidal disruption of dSph satellites in the Galactic halo
and, in particular, to search for new tidal tails in a sample of nearby dSph
satellites of the Milky Way. Here we present our finding of extra-tidal debris in the Ursa Minor dSph and discuss the detection by the Sloan Digitized Sky
Survey team of what could be a tidal stream associated to the Sagittarius dSph.
\end{abstract}

\end{opening}           

\section{Introduction}  

The
formation of the Galactic halo is currently best explained by the
combination of two scenarios which were previously  regarded as
competing models. Based on the kinematics of metal-poor halo field
stars, Eggen, Lynden-Bell \& Sandage (1962) proposed that the
halo formed during a rapid, smooth collapse from a homogeneous
primordial medium. Searle \& Zinn (SZ, 1978) argued 
halo formation via the gradual merging of many sub-galactic
fragments. 

Although recent evidence shows that the inner region of the Galactic halo ($R<20 kpc$) has mostly been formed in a fast process (Rosenberg {\it et al.} 1999), several results indicate at least a part of the outer halo originated in a process similar to the SZ scenario. The discovery of the Sagittarius
dwarf galaxy (Ibata, Gilmore \& Irwin 1994), in the process of
dissolving into the Galactic halo, argued in favour of the hypothesis
that accretion
events can take place in the Milky Way, whose  full formation
history (through satellites merging into it) might not have  finished
yet. On the basis of this discussion, it is very
important to investigate whether Galactic  dSph
satellites display tidal tails beyond their tidal limits. The availability of a new generation of wide-field CCD
cameras offer for the first time a good opportunity of successfully addressing this issue.

\section{Methodology}

The detection of tidal tails in dSphs is very challenging due to
their large angular sizes and low surface brightness. This requires using
wide field observations and a careful analysis of the foreground
contamination.  We use a technique based in B,R photometry survey of selected wide fields of the galaxy to obtain deep 
color-magnitude diagrams (CMDs) that reveal the  main sequence (MS) turnoff of its old population. In moderate foreground contaminated fields, it is also possible to trace the tidal debris by means of the blue horizontal stars (BHB) or blue straggles
(BS) members of the dSph, due to the almost absence of Galactic foreground stars for $(B-R)< 0.5$.

\section{Results}

\subsection{Ursa Minor}

Ursa Minor (UMi) is one of the closest satellites of the Milky Way (d=69
kpc) and a strong candidate to be a disrupted dSph interacting with the external
Galactic halo. Figure 1 shows the $[(B-R),V]$ CMD for three selected elliptical annuli 
centered in UMi: a) the central region; b) the
elliptical area beyond the tidal radius ($r_{t}$) given by Kleyna {\it et al.} (1995) ($R=34'$); and c)
the elliptical region beyond the $r_{t}$ obtained by Irwin \& Haztzidimitriou (IH, 1995) ($R=50.6'$). For comparison, the CMD of a control field situated $\sim 3 \deg$ S from the center of UMi is shown in Figure 1d. BHB as well as old MS turnoff stars are detected in these extra-tidal fields (Figure 1b and 1c) indicating the presence of a 
tidal extension in Ursa Minor even beyond the $r_{t}$ given by IH.

The existence of tidal tails in UMi suggests that this satellite
is undergoing a tidal destruction process. This is also supported by the presence of  
substructure in the main body of UMi reported by Olszewski \& Aaronson (1985) and more recently by IH (1995) and Kleyna {\it et al.} (1998). We confirm this lumpiness and asymmetry in the stellar distribution of UMi from our deeper data, although we are currently carrying on an analysis to test its statistical significance. In this context, it is possible a tidal origin for the  UMi's  high observed mass-to-light ratio, as it is suggested
by Kroupa (1997). If this substructure is real, more 
elaborated models including details of the substructure and the presence
of tides will be needed to estimate the real dark matter content of UMi.

\begin{figure}
\centerline{\includegraphics[width=20pc]{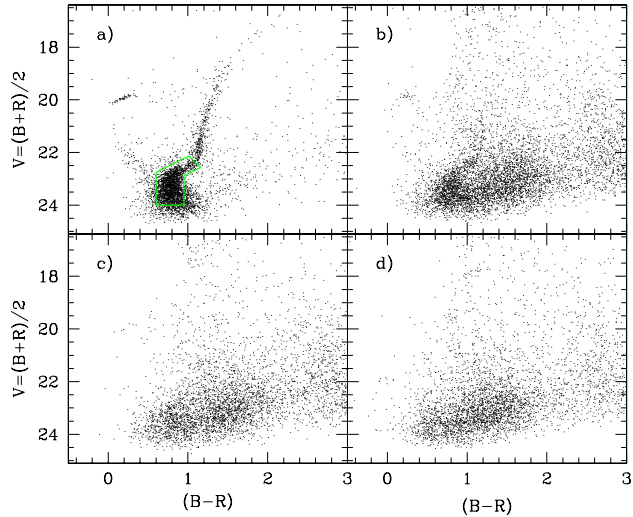}}
\caption{CMDs for the UMi regions and a control field: a) $R<6.6'$; b) $34' < a < 50.6'$; c) $R> 50.6'$; d) the control field.}

\end{figure}

\subsection{Sagittarius}

There is general agreement that the Sagittarius (Sgr) satellite is being
disrupted by the Milky Way. Theoretical simulations
of the encounter (G\'omez-Flechoso {\it et al.} 1999) predict tidal streams emerging
from the main body of Sgr and extending along its projected major axis, 
and possibly even encircling the sky.

Recently, the Sloan Digital Sky Survey (SDSS) have found two clear, $\sim$ 45 deg long stripes of blue, A-type stars, 
with magnitudes 19 and 21 (Yanni {\it et al.} 2000) . They could be respectively formed by BHB and BS stars at 45 kpc from the Sun belonging to an 
old stream in the outer galactic halo, possibly associated to a tidally 
disrupted dwarf galaxy. 

The best of all known candidates is Sgr, due to the SDSS slice overlaps the area where the models predict the presence of the Sgr northern stream.
To check this possibility we have computed a model of Sgr assuming that the two streams found by the SDDS are tidal debris of this galaxy. The result is shown in Figure 2. The agreement is very good (see Figure 3 in Yanni et al. 2000)  and suggests we are likely seeing the apocenter of Sgr, although  the possibility of an  unknown tidal disrupted galaxy cannot be rejected. We
are carrying on a photometry survey in this region to spatially trace the
stream.

\begin{figure}
\centerline{\includegraphics[width=20pc]{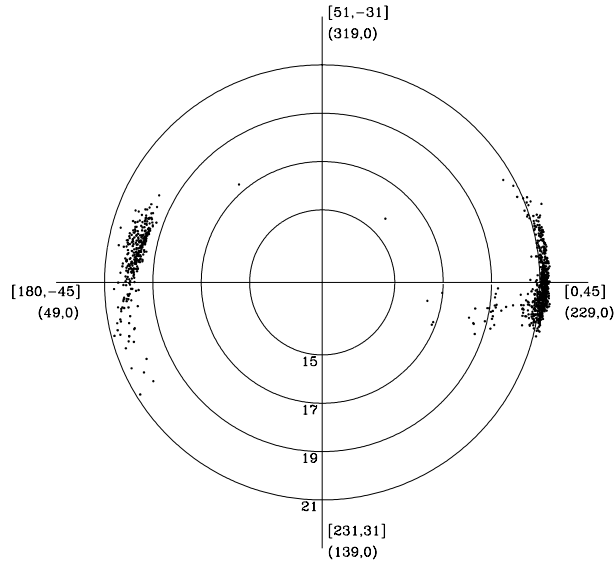}}
\caption{A polar wedge diagram with Right Ascension and $V$ magnitude for
our model of Sgr (see Yanni {\it et al.} 2000 for details of this representation).}
\end{figure}

\theendnotes

\end{article}
\end{document}